# Evaluating the Design Features of an Intelligent Tutoring System for Advanced Mathematics Learning

Ying Fang[1,2] Bo He[1,2], Zhi Liu[1,2], Sannyuya Liu[1,2], Zhonghua Yan[1,2], Jianwen Sun[1,2*]

[1]Faculty of Artificial Intelligence in Education, Central China Normal University, Wuhan 430079, China
[2]National Engineering Research Center of Educational Big Data, Central China Normal University, Wuhan 430079, China
`{fangying,hb925822,zhiliu,liusy027,hit_yan,sunjw}@ccnu.edu.cn`

**Abstract.** Xiaomai is an intelligent tutoring system (ITS) designed to help Chinese college students in learning advanced mathematics and preparing for the graduate school math entrance exam. This study investigates two distinctive features within Xiaomai: the incorporation of free-response questions with automatic feedback and the metacognitive element of reflecting on self-made errors. An experiment was conducted to evaluate the impact of these features on mathematics learning. One hundred and twenty college students were recruited and randomly assigned to four conditions: (1) multiple-choice questions without reflection, (2) multiple-choice questions with reflection, (3) free-response questions without reflection, and (4) free-response questions with reflection. Students in the multiple-choice conditions demonstrated better practice performance and learning outcomes compared to their counterparts in the free-response conditions. Additionally, the incorporation of error reflection did not yield a significant impact on students' practice performance or learning outcomes. These findings indicate that current design of free-response questions and the metacognitive feature of error reflection do not enhance the efficacy of the math ITS. This study highlights the need for redesign or enhancement of Xiaomai to optimize its effectiveness in facilitating advanced mathematics learning.

**Keywords:** Intelligent Tutoring System, Mathematics, Item Format, Reflection

## 1 Introduction

In recent years, there has been a significant increase in the number of students enrolling in graduate school entrance exams in China. The enrollment figure rose from 3.41 million in 2020 to 4.74 million in 2023, with an average annual growth rate of 15.8% over the past seven years [1]. While advanced mathematics is generally considered challenging by college students, for those in science and engineering fields, passing the graduate-level mathematics entrance exam is a necessary requirement for admission to postgraduate studies [2,3]. To address this, we developed an intelligent tutoring system (ITS) called Xiaomai, aimed at helping students learn advanced mathematics (i.e., college-level mathematics) and prepare for the graduate school math entrance exam.

The system has a question bank of nearly 5,000 mathematics problems which were carefully selected by domain experts, targeting 177 college-level mathematics knowledge components. One distinctive characteristic of the system is the integration of free-response questions, a feature rarely implemented in math ITSs due to the challenge of automated evaluation. Xiaomai tackles this by providing correct solutions upon answer submission, and then requiring students to self-evaluate their answers. Additionally, the system includes an error reflection feature that prompts students to reflect on the rationale behind errors when mistakes are made, with the goal of helping students monitor their learning. This study focuses on these two characteristics of Xiaomai, aiming to assess their impacts on promoting students' mathematics learning. The findings of the study can inform the current design of the system and guide future improvements.

## 2       Background and Related Work

### 2.1       Xiaomai - an ITS for Advanced Mathematics with Metacognitive Features

Xiaomai consists of instruction, practice and assessment modules covering 177 knowledge components (KCs) in linear algebra, calculus and statistics. The system organizes all KCs into a comprehensive domain knowledge tree, with KCs being the leaf nodes, and mapping relationships established between questions and KCs. Additionally, a graph indicating the prerequisite relations of knowledge components was created to facilitate effective learning path navigation. The categorization of KCs, questions, and the order of the KCs are all derived from domain experts' annotations.

The KCs and questions are recommended to students through a hybrid recommender system [4] implementing rule-based and knowledge tracing algorithms. Specifically, the system applies a set of rules to obtain candidate subsets, and then selects the best target with a collaborative filtering recommendation method. The rule-based filtering can reduce the computation burden of the subsequent collaborative filtering by narrowing the search scope, thus improving the recommendation efficiency. Figure 1 illustrates the process and the rules designed for selecting KCs.

   a. **Novelty Enhance:** Recommend KCs that have not been previously studied to enhance the novelty of the recommendation system and ensures a comprehensive coverage of all knowledge components.

   b. **Drawback Discover:** Recommend KCs with higher error rates, as these KCs are more likely to reveal cognitive deficiencies.

   c. **Forgetting Mitigate:** Recommend KCs that have not been reviewed for a preset period to mitigate the risk of forgetting.

   d. **Cognitive Prefer:** Recommend KCs that learners have frequently selected autonomously, as these choices can reflect certain cognitive preferences.

   e. **Cognitive Demand:** Recommend neighbors of the current KC on the prerequisite graph, where predecessors serve as an essential foundation for acquiring the current KC, while successors guide the future learning direction.

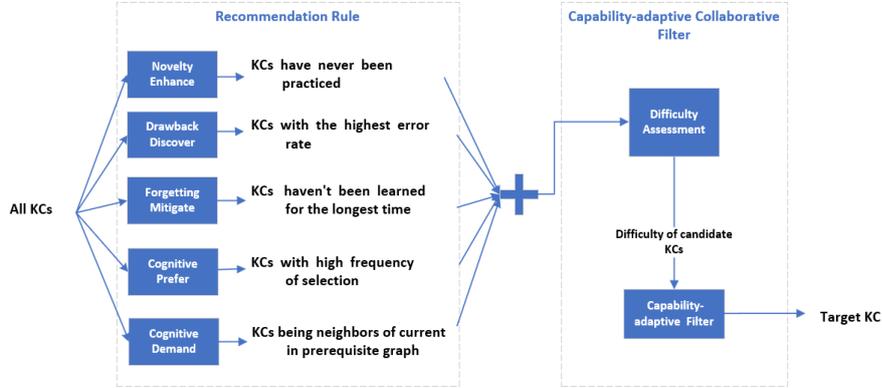

**Fig. 1.** Knowledge components recommendation mechanism in Xiaomai.

After obtaining a preliminary set of candidate KCs using the aforementioned rules, the most suitable KC is selected through the capability-adaptive collaborative filter. The filter considers the discrepancy between the difficulty level of the KCs and the student's mastery level, which is inferred by knowledge tracing models. The KC that is best aligned with the student's current knowledge state is selected for recommendation. Question recommendation follows a similar procedure.

Automatic feedback is generated upon answer submission, displaying answer correctness, detailed solutions, KCs related to the question, and response time. When an error occurs, a prompt is generated by the system, typically providing the step-by-step solution and recommending KCs to learn.

In the chapter assessment module, students are provided with a set of questions related to the KCs covered in a chapter to diagnose their mastery. A comprehensive report is generated at the end of the assessment, as is shown in Figure 2.

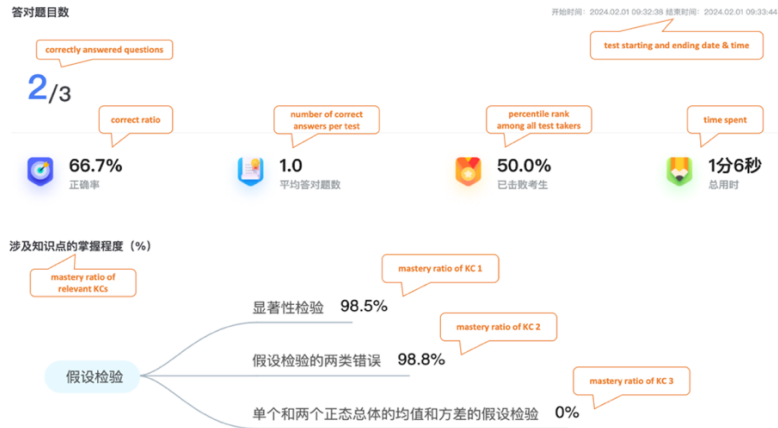

**Fig. 2.** A chapter assessment report in Xiaomai.

Xiaomai incorporates free-response questions that prompt students to upload step-by-step images of their answers into the system. As shown in Figure 3, when students click on the square labeled "upload problem-solving process," a QR code appears on the screen, allowing them to scan and upload up to three pictures to the system. Upon submitting their answers, a detailed solution is displayed, and students are prompted to self-evaluate their responses using the provided scale.

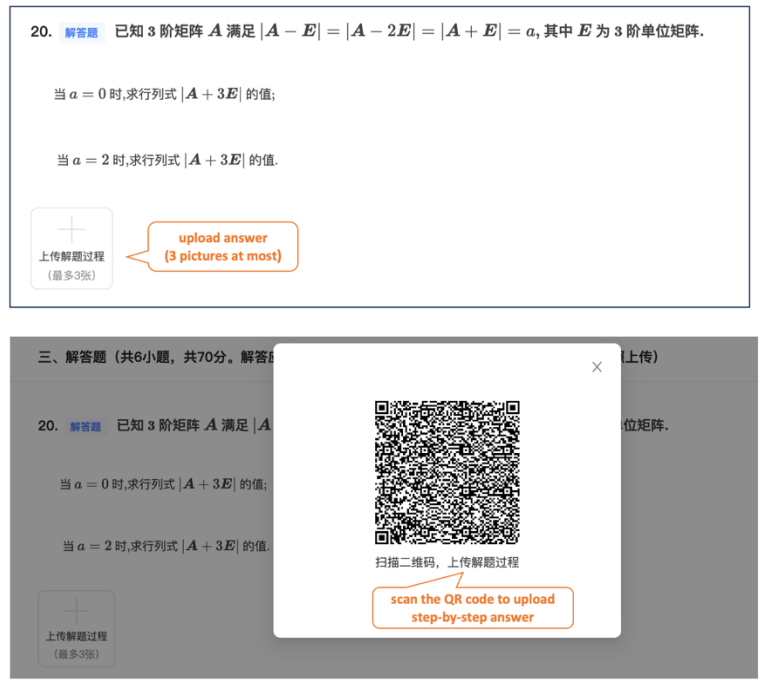

**Fig. 3.** A QR code is displayed after clicking "upload answer" area for students to upload their answer images for the free-response question.

Xiaomai is designed for college students, who are assumed to have some metacognitive skills. As such, metacognitive elements have been integrated into Xiaomai, allowing students to monitor, regulate, and reflect on their learning. The interface of Xiaomai includes a "History" tab where students' learning and practice details are recorded. This learning history can serve as a reminder of students' progress and assist them in planning their future learning. Additionally, there is an "Error Collection" tab that automatically records mistaken questions, and a "Bookmark" tab that collects questions bookmarked by students themselves. When viewing mistaken questions, students have the option to view by chapter or by the cause of mistakes. Viewing by chapter may help students identify their weaknesses in particular chapters, while viewing by the cause of mistakes may assist students in understanding how to avoid specific errors in the future. Despite the automatic generation of the Error Collection, students can remove questions from the collection, making it easier for them to track the KCs they haven't mastered yet.

In addition to the tabs mentioned above, the features of self-evaluation and error reflection require students' engagement in metacognition. For free-response questions, students can write their answers on paper, take a picture, and then scan the QR code on the screen to upload their answers. Upon submission, step-by-step solutions are displayed, prompting students to self-assess their answers using the provided binary or categorical scales. When an error occurs (i.e., students' self-evaluated score is not perfect), students are asked to select the reason for the error from a provided list (i.e., misunderstand the question, calculation error, lack knowledge of the KC, forget the KC, carelessness, lack problem-solving strategy) or report it if the reason is not listed.

## 2.2 Free-response versus Multiple-choice Questions for Learning

Both free-response and multiple-choice questions are commonly used to facilitate and assess student learning. Multiple-choice questions are efficient for grading, providing an easy-to-use assessment tool. In contrast, free-response questions offer insights into students' thought processes and can help teachers identify misconceptions and evaluate various techniques students use [5]. Prior research has explored the impact of the two question formats on assessment and learning outcomes, resulting in mixed findings.

Regarding assessment, free-response questions were found to yield more information and provide more accurate estimates of students' science ability, although their advantage over multiple-choice questions was considered to be relatively small [6,7]. In contrast, carefully designed multiple-choice questions were found to accurately reflect free-response question performance while maintaining the ease of grading and quantitative analysis [8]. In math assessment, multiple-choice questions were suggested as providing less diagnostic information than free-response formats, limiting their ability to accurately measure problem-solving skills [9].

In addition to assessing students' knowledge and skills, both multiple-choice and free-response questions are used during instruction to facilitate learning. Larsen et al. reported that free-response questions improved students' performance on subsequent free-response questions in a traditional class setting [10]. McDaniel et al. found that the effect sizes for short-answer and multiple-choice questions on students' subsequent performance on the same type of questions in an online course were similar [11]. Roediger et al. demonstrated that multiple-choice questions improved performance not only on subsequent multiple-choice questions but also on short-answer questions [12]. Glass and Sinha reported that multiple-choice questioning during instruction effectively improved exam performance in middle-school and college classes, regardless of whether it was done in class or online [13]. Kang et al. found that free-response questions led to better results in the posttest compared to multiple-choice questions when students were provided with immediate feedback during their intervention [14]. These studies focus on learning in the domains of psychology, social science, and reading. Regarding mathematics, Stankous reviewed previous studies comparing multiple-choice and constructive-response formats for math learning and concluded that constructive-response questions promote student learning better than multiple-choice questions in mathematics [15].

### 2.3 Effect of Error Reflection on Learning

Reflection on errors is widely recognized as promoting learning and knowledge acquisition [16,17]. Previous research emphasize that learners develop more comprehensive cognitive models by reflecting on both correct and incorrect facts, conceptions, and strategies [18,19]. Studies on self-regulated learning suggest that reflecting on the reasons for errors can enhance students' understanding of their own cognitive functioning and thus facilitate their learning [20,21].

In the mathematics domain, learners who explained both correct and incorrect examples demonstrated better performance compared to those who only explained correct examples [17, 22, 23]. Heemsoth and Heinze found that reflecting on the rationale behind self-made errors in fractions practice enhanced students' procedural and conceptual knowledge [24]. In addition, some research found that the impact of reflection was moderated by prior knowledge. Specifically, students with high prior knowledge students benefited from error reflection, whereas students with little prior knowledge mainly benefited from reflecting on correct examples [16, 25]. However, Durkin and Rittle-Johnson found the interaction effect between prior knowledge and reflection on learning was not significant [23].

### 2.4 Current Study

The goal of this study is to assess the efficacy of two design features in the math ITS known as Xiaomai: the incorporation of free-response questions and the metacognitive aspect of error reflection. Specifically, students are required to write down the step-by-step answers for the free-response questions. Upon submission, students receive detailed solutions and are prompted to self-assess their own answers. Error reflection is triggered by an incorrect answer, and students must provide the reason for the error. This study investigates the impact of free-response questions compared to the commonly implemented multiple-choice questions in math ITSs. Additionally, it examines whether reflection on errors promotes more effective math learning. The following research questions are addressed in this study:

1. To what extent do item format and error reflection affect learning outcomes of advanced math?
2. To what extent do item format and error reflection influence students' performance in an ITS for advanced math learning?

## 3 Methods

### 3.1 Design and Participants

Participants were 129 students recruited from a university in central China. All participants reported taking advanced mathematics courses in college during the screening process. Nine participants were excluded from the study due to incomplete participation or failing the attention check, resulting in a final sample of 120 (45 male, 75 female) with an average age of 20.13 years. Participants were given financial compensation of ¥60 for their participation.

The study utilized a 2 by 2 factorial design with item format (multiple-choice questions vs. free-response questions) and reflection (reflection vs. no reflection) as independent variables. Participants were randomly assigned to one of the following conditions: (1) Multiple-choice Questions without Reflection, (2) Multiple-choice Questions with Reflection, (3) Free-response Questions without Reflection, and (4) Free-response Questions with Reflection. This design allowed for the examination of the effects of different question formats and the presence of reflection on participants' performance and learning outcomes.

### 3.2 Procedure, Materials and Measures

Participants first completed a 20-minute pretest consisting of 18 multiple-choice questions. They were then randomly assigned to one of the four experimental conditions to study on three knowledge components by solving the related math problems in Xiaomai for 40 minutes.

In the Multiple-choice Questions without Reflection condition, all the questions were in multiple-choice format with immediate feedback showing the correctness of the answers and detailed solutions. In the Multiple-choice Questions with Reflection condition, participants received the same questions and feedback but were prompted to reflect and report the reasons for the error when an error occurs. In the Free-response Questions without Reflection condition, participants were provided with multiple-choice and free-response questions alternatively, with the latter requiring step-by-step answers and then self-rating their responses based on the provided solutions. In the Free-response Questions with Reflection condition, participants received the same questions and feedback as the previous condition, and were also prompted to reflect on their errors and report the reasons for the errors. The practice questions were selected and edited by a domain expert to ensure content equivalency.

After the practice session, participants completed a 20-minute posttest containing 18 multiple-choice questions. A domain expert selected the 18 questions for the pretest and posttest from Xiaomai question bank, ensuring that the questions in the posttest covered the same knowledge components and were at the same difficulty levels as their counterparts in the pretest.

**Pretest and Posttest.** The outcome of pretest and posttest were measured by the number of correct answers.

**Practice performance.** Participants' performance measures included the number of attempts, answer correctness, and time spent on the questions. For multiple-choice questions, answer correctness was recorded as 1 for correct and 0 for incorrect. Regarding free-response questions, an answer was recorded as 1 when self-rated as "completely correct" or achieving the full mark for the given measure. Time was measured in seconds.

### 3.3 Data Analyses

One goal of this study is to investigate the effects of item format and reflection on learning outcomes, measured as the pretest to posttest change. A two-way ANOVA analysis was conducted to analyze the effect of item format and reflection on posttest scores while controlling for pretest scores. Pretest to posttest effect size (Cohen's d) for different item format and reflection groups were also computed.

Another goal is to examine the effects of item format and reflection on participants' performance during their practice in Xiaomai. Two t-tests were conducted to examine the effect of item format and reflection on the number of attempts during practice. Mixed-effects linear regressions were performed to analyze the effect of item format and reflection on item correctness and time per item using the lme4 package in R (Bates et al., 2014), with subjects specified as a random factor to adjust for subject variance.

## 4 Results

### 4.1 Descriptive Statistics of Test scores and Practice Performance

Table 1 shows the means and standard deviations of the pretest and posttest scores, as well as the raw score sand the proportion of correct answers during practice for each condition. One-way ANOVAs (analysis of variance) were conducted to compare the means of test scores and practice performance between the four conditions. The results indicated no statistically significant difference between the four conditions regarding pretest score ($F(3, 116) = 0.23$, $p = 0.88$), posttest score ($F(3, 116) = 0.76$, $p = 0.52$), or practice proportion correct ($F(3, 116) = 1.43$, $p = 0.24$). Practice performance scores showed significant difference between the four conditions ($F(3, 116) = 5.95$, $p = 0.01$). Subsequent post hoc tests using Tukey's HSD revealed that the scores in the two multiple-choice questions conditions were significantly higher than those in the two free-response questions conditions.

**Table 1.** Means and standard deviations of pretest, posttest and practice performance scores.

| Condition | N | Pretest M (SD) | Posttest M (SD) | Practice Score M (SD) | Proportion Correct M (SD) |
|---|---|---|---|---|---|
| MC | 33 | 6.15 (2.12) | 8.88 (3.94) | 13.15 (5.95) | 0.59 (0.17) |
| MC-R | 31 | 6.55 (2.58) | 9.35 (3.34) | 12.94 (6.38) | 0.61 (0.15) |
| FR | 28 | 6.68 (3.95) | 8.00 (4.96) | 8.36 (5.93) | 0.55 (0.19) |
| FR-R | 28 | 6.64 (2.51) | 8.11 (3.82) | 8.46 (5.71) | 0.52 (0.16) |

Note: MC = Multiple-choice question without reflection, MC-R = Multiple-choice question with reflection, RF = Free-response question without reflection, RF-R = Free-response question with reflection.

### 4.2 The Impact of Item Format and Reflection on Learning Outcome

A Two-way ANCOVA was conducted to examine the effect of item format, reflection and their interaction on posttest scores while controlling for pretest scores. The results indicated that there was no statistically significant interaction effect between item format and reflection on posttest scores ($F(1,115) = 0.00$, $p = 0.95$). The main effect of item format was marginally significant ($F(1,115) = 3.72$, $p = 0.06$), while the main effect of reflection was not statistically significant ($F(1,115) = 0.07$, $p = 0.79$). As is shown in Table 2, the participants in the multiple-choice conditions ($M = 9.21$) achieved higher posttest scores compared to the participants in the free-response conditions ($M = 7.94$) after controlling for their pretest scores. However, the reflection did not appear to affect participants' posttest scores after controlling for pretest scores. Additionally, we computed the effect sizes (Cohen's d) of the four groups based on the mean pretest-to-posttest score changes, as shown in Figure 4. The effect sizes were large for the multiple-choice question conditions, yet small for the free-response question conditions. Meanwhile, the effect sizes for both no reflection and reflection conditions were medium.

**Table 2.** Estimated marginal means and standard errors of posttest scores for different conditions.

| Main-effect Variable | Condition | N | Mean | Std. Error | 95% CI |
|---|---|---|---|---|---|
| Item Format | Multiple-choice Question | 64 | 9.21 | 0.45 | [8.32, 10.10] |
| | Free-response Question | 56 | 7.94 | 0.48 | [6.99, 8.89] |
| Reflection | No Reflection | 61 | 8.49 | 0.46 | [7.58, 9.40] |
| | Reflection | 59 | 8.66 | 0.47 | [7.74, 9,59] |

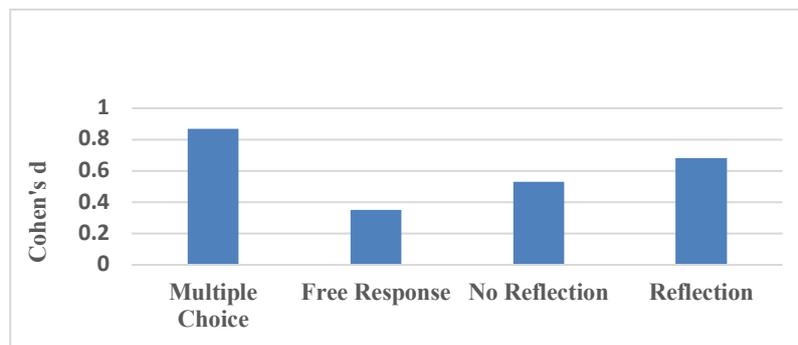

**Fig. 4.** Effect sizes of different item format and reflection conditions.

Given that prior research suggested that only students with high prior knowledge may benefit from error reflection, we categorized the participants into high and low prior knowledge groups using a median split. An ANOVA was conducted to assess the impact of prior knowledge, reflection, and their interaction on pretest-to-posttest learning gain. Neither the main effect of reflection nor the interaction effect was found to be statistically significant.

**4.3 The Impact of Item Format and Reflection on Practice Performance**

Two t-tests were performed to examine the effect of item format and reflection on the number of attempts during practice. Results revealed that participants in the multiple-choice conditions ($M = 21.98$, $SD = 8.11$) attempted significantly more questions ($t(118) = 4.29$, $p < .001$) than participants in free-response conditions ($M = 15.41$, $SD = 8.61$). However, there was no statistically significant difference ($t(118) = 0.47$, $p = 0.64$) in the number of attempts between conditions with reflection ($M = 18.53$, $SD = 8.76$) and without reflection ($M = 19.30$, $SD = 9.17$).

Two mixed-effects linear regressions were conducted to examine the impact of item format and reflection on the correctness of answers. In both models, subjects were included as a random-effect factor to account for subject variance. In the model predicting item correctness with item format, the fixed effect was marginally significant ($p = 0.09$), indicating that multiple-choice questions were more likely to be answered correctly. The model predicting item correctness with reflection found the fixed effect was not nonsignificant ($p = 0.81$) (see Table 3).

**Table 3.** Fixed effects of item format and reflection on answer correctness and response time.

| Fixed-effect Variable | Predicted Variable | B | Std. Error | t | p |
|---|---|---|---|---|---|
| Item Format | Correctness | 0.23 | 0.14 | 1.67 | 0.09 |
| Reflection | Correctness | 0.03 | 0.13 | 0.24 | 0.81 |
| Item Format | Time | 46.26 | 8.92 | 5.19 | < .001 |
| Reflection | Time | -3.74 | 9.82 | 0.38 | 0.70 |

Two mixed-effects linear regression analyses were performed to explore the impact of item format and reflection on time per question. In both models, subjects were included as a random-effect factor to account for subject variance. The model predicting time per question based on item format revealed a statistically significant fixed effect ($p < 0.001$), indicating that participants spent less time on multiple-choice questions compared to free-response questions. However, the model predicting time per question based on reflection found that the effect was not statistically significant ($p = 0.70$). These results are shown in Table 3.

## 5    Discussion

We developed an ITS called Xiaomai to assist Chinese college students in mastering advanced mathematics and preparing for the graduate school math entrance exam. Given that free-response questions accounted for approximately half of the total score on the math entrance exam, we incorporated this question format into Xiaomai. However, automatically evaluating free-response answers in mathematics presents a significant challenge. To address this issue, our approach involves providing detailed correct solutions upon answer submission and allowing students to self-assess their responses. Furthermore, we integrated an error reflection feature into the system to support students in monitoring their progress. In this paper, we reported an experiment conducted to assess the effectiveness of free-response questions and error reflection in facilitating students' math learning.

In terms of item format, the results of our analyses indicated that students performed better during practice and achieved greater learning gains when the math problems were in a multiple-choice format compared to a free-response format. It is important to note that scores for free-response questions were only recorded as "1" if students self-evaluated them as completely correct. Therefore, if a student obtained a correct result but missed some steps, it might be scored as "1" for a question in multiple-choice format, but as "0" for a free-response question. The evaluation criteria were more stringent for free-response questions than for multiple-choice questions to some extent. Additionally, the questions in both the pretests and posttest were in multiple-choice format. Previous research has shown similar effect sizes for multiple-choice and free-response questions regarding subsequent performance on the same type of questions [11]. Complementary to prior research, our study revealed that practicing with questions in a different format (i.e., free-response format) was not as effective as practicing with questions in the same format for improving students' subsequent performance on multiple-choice questions.

In terms of error reflection, our findings indicated that reflecting on the reasons behind errors did not contribute to performance in Xiaomai or learning gains, which contradicted our expectations. Although reflection is mandatory when an error occurs, there is no mechanism in the system to ensure students engage in reflection seriously. It is possible they select a reason randomly just to move to the next question. Response time for each question is recorded in Xiaomai, but the reflection time is not recorded separately. Thus, we cannot infer students' engagement using the time they spent on reflection, which is the limitation of this study. To better understand why reflection was ineffective, additional information on students' reflection process needs to be collected.

We also investigated students' prior knowledge, as previous studies yielded mixed results on whether the impact of reflection is influenced by prior knowledge. In line with Durkin and Rittle-Johnson's findings [23], our results suggested that the impact of reflection was not influenced by prior knowledge.

The findings of this study emphasize the need to redesign or enhance Xiaomai to optimize its effectiveness in facilitating advanced mathematics learning. For example, in addition to correct step-by-step solutions, students could be presented with typical incorrect solutions and related misconceptions for free-response questions. We may also consider reducing the proportion of free-response questions in Xiaomai given the evidence that multiple-choice format is more effective in promoting student learning. In addition, we need to collect data on the process of student reflection to determine whether to retain this feature or how to improve it so that students may effectively monitor their learning through reflection.

Evaluating the Design Features of an ITS for Advanced Mathematics

**Acknowledgement.** This research is financially supported by the National Science and Technology Major Project (2022ZD0117100), National Natural Science Foundation of China (61937001), and Hubei Provincial Natural Science Foundation of China (2023AFA020). Any opinions, findings, and conclusions or recommendations expressed in this material are those of the authors and do not necessarily reflect the funders.